\documentclass{aa}
\usepackage{graphicx}
\usepackage{txfonts}
\usepackage{epsfig}
\usepackage{times}
\def\ltsima{$\; \buildrel < \over \sim \;$}
\def\gtsima{$\; \buildrel > \over \sim \;$}
\def\simlt{\lower.5ex\hbox{\ltsima}}

\def\simgt{\lower.5ex\hbox{\gtsima}}

\begin{document}

   \title{X-ray observations of the Compton-thick Seyfert 2 galaxy, NGC 5643}

   \author{G. Matt \inst{1},  S. Bianchi\inst{1}, A. Marinucci\inst{1}, M. Guainazzi\inst{2}, 
K. Iwawasa\inst{3}, E. Jimenez Bailon\inst{4}\\ 
}
\authorrunning{G. Matt et al. }
\titlerunning{X-ray observations of NGC 5643}

   \offprints{G. Matt, \email{matt@fis.uniroma3.it} }

   \institute {$^1$Dipartimento di Matematica e Fisica, Universit\`a degli Studi Roma Tre, 
via della Vasca Navale 84, I-00146 Roma, Italy \\
$^2$ European Space Astronomy Center of ESA, Apartado 50727, 28080 Madrid, Spain \\
$^3$  ICREA and Institut de Ciències del Cosmos (ICC), Universitat de Barcelona (IEEC-UB), 
Martí i Franquès 1, 08028, Barcelona, Spain \\
$^4$ Universidad Nacional Autonoma de Mexico, Apartado Postal 70 - 264
Ciudad Universitaria,  D.F., CP 04510, Mexico\\
}

   \date{Received / Accepted }

   \abstract{We present results from a $\sim$55 ks long XMM-$Newton$ observation of the 
obscured AGN, NGC 5643, performed in July 2009. A previous, shorter (about 10 ks)
XMM-$Newton$ observation in February 2003 had left two major
issues open,  the nature of the hard X-ray emission (Compton--thin vs Compton--thick)
and of the soft X-ray excess (photoionized vs collisionally ionized matter).
The new observation shows that the source is Compton--thick and that the
dominant contribution to the soft X-ray emission is  
by photoionized matter (even if it is still 
unclear whether collisionally ionized matter may contribute as well). \\
We also studied three bright X-ray sources that are in the field
of NGC 5643. The ULX NGC 5643 X-1 was confirmed to be very luminous, even if more 
than a factor 2 fainter than in 2003. We then provided the first high quality 
spectrum of the cluster of galaxies Abell 3602. The last source,
CXOJ143244.5-442020, is likely an unobscured AGN, possibly belonging to Abell 3602. }
   \maketitle

%

\section{Introduction}

 NGC 5643 is a nearby (z=0.004) SAB(rs)C galaxy, 
known to host a low-luminosity Seyfert 2 
nucleus (Phillips et al. 1983). An extended emission-line region 
elongated in a direction close to the radio position angle (Morris et al. 1985),  
is probably the projection of a 1.8-kpc, 
one-sided ionization cone (Simpson et al. 1997). NGC 5643 belongs to the class 
of `extreme infrared' (IR) galaxies (Antonucci \& Olszewski 1985).
Although intense episodes of star formation are occurring in a nearly circular arm, 
mid-IR diagnostics suggest that the AGN dominates the IR 
energy budget (Genzel et al. 1998). Comparisons of optical spectra with synthesis models are, 
however, consistent with a `starburst/Seyfert 2 composite' spectrum (Cid Fernandes et 
al. 2001). The NGC 5643 nucleus is a strong radio emitter as well, most likely powered 
by the AGN (Kewley et al. 2000). 

In X-rays, NGC 5643 has been observed by ASCA, BeppoSAX, $Chandra$ and XMM--$Newton$. 
The $Chandra$ image shows that the soft X--ray emission is spatially correlated 
with the HST [O III] emission, like in many Seyfert 2s (Bianchi et al. 2006).
The extended component accounts for  about half of the total soft X--ray emission.
The soft X-ray/[O III] flux ratio is similar to that of the other sources of the 
Bianchi et al. (2006) sample, suggesting a common origin in a photoionized medium. 
On the other hand,  the line 
diagnostics derived from the XMM-Newton RGS spectrum failed to reach a definitive 
answer on the dominant ionization process at work in the gas (Guainazzi \& Bianchi 2007).
The XMM-$Newton$ hard X-ray spectrum is very flat, but whether this is due to 
Compton-thin absorption or to reflection from neutral matter 
(indicative of a Compton-thick source) cannot be said from that short (less
than 10 ks) observation (Guainazzi et al. 2004).

Interestingly, when the XMM--$Newton$ spectrum is
compared with previous BeppoSAX and ASCA observations, dramatic
spectral variability is found. However, the PSF of the X--ray telescopes of these
two satellites were not good enough to separate emission of the nucleus of 
NGC 5643 from that of 
a nearby (0.8' away) X-ray source (christened NGC 5643 X--1 by Guainazzi et al. 2004) which,
at the time of the XMM--$Newton$ observation, was about as bright as the nucleus of NGC 5643 in the
2--10 keV band, and twice as bright in the 0.5-2 keV range. It is therefore impossible
to ascribe the variability to either the NGC 5643 nucleus or X--1. It is worth noting that, 
if indeed belonging to NGC 5643, X--1 would be one of the brightest 
ULX known, being therefore interesting on its own.  

In this paper we analyse a longer XMM-$Newton$ observation, with the double aim of
checking whether the soft X-ray emission is due to photoionized or 
collisionally ionized plasma
and to determine whether the source is Compton-thin or Compton-thick. We also 
analyse NGC 5643 X--1 and 
the other bright sources present in the field.

The paper is organized as follows. In Sec.2 the data reduction is described, while
data analysis and results on NGC 5643 
and on the brightest sources in the field
(the ULX NGC 5643 X-1, the cluster of galaxies Abell 3602, the source 
CXOJ143244.5-442020 which is possibly an active galaxy belonging to Abell 3602) are
reported in Sec.3 and Sec.4, respectively. The main results of the work are summarized
in Sec.5.

In the following, the standard cosmological model ($H_0$=70 km/s/Mpc, $\Omega_M$=0.3 and
$\Lambda$=0.7) will be assumed.

\section{Data reduction}

NGC~5643 was observed by XMM-\textit{Newton} twice, for $\simeq10$ ks on 2003-02-08 
(\textsc{obsid} 0140950101) and $\simeq55$ ks on 2009-07-25 (\textsc{obsid} 0601420101). 
In both cases, the observations were performed with the EPIC CCD cameras, the pn and the 
two MOS, operated in full frame, and medium filter. Data were reduced with SAS 11.0.0 and 
screening for intervals of flaring particle background was done consistently
with the choice of extraction radii, in an iterative process based on the procedure to
maximize the signal-to-noise ratio described by Piconcelli et al. (2004). After this 
process, the net exposure times were of about 7 (45), 9 (53), and 9 (51) ks for pn, MOS1 
and MOS2, respectively, in \textsc{obsid} 0140950101 (0601420101), adopting extraction 
radii of 25 arcsec for all the cameras in the case of the nucleus of NGC5643. Extraction 
radii for NGC5643 X-1 are 26 arcsec for all the cameras, 21 arcsec for CXOJ143244.5-442020, 
and 90 arcsec for Abell 3602. The background spectra were extracted from source-free circular 
regions with a radius of 50 arcsec. Patterns 0 to 4 were used for the pn spectrum, while MOS 
spectra include patterns 0 to 12. Since the two MOS cameras were operated with the same mode, 
we co-added MOS1 and MOS2 spectra, after having verified that they agree with each other and 
with the summed spectrum. Finally, spectra were binned in order to oversample the instrumental 
resolution by at least a factor of 3 and to have no fewer than 30 counts in each 
background-subtracted spectral channel. The latter requirement allows us to use 
the $\chi^2$ statistics. Spectra were analysed with XSPECv12.6.0.

RGS spectra were extracted from event lists reduced
through the SAS meta-task {\tt rgsproc}. We used the optical
coordinates of NGC~5643 from NED to fix the reference of the
wavelength scale. In both XMM-Newton observations, NGC~5643 X-1 was
placed along a direction perpendicular to the dispersion direction
with respect to the AGN. This allowed us to minimize the
contamination by NGC~5643 X-1 in the AGN spectrum by restricting the
extraction region of the latter to 50\% of the cross-dispersion
point spread function (the exact number was chosen to maximize the
signal-to-noise of the AGN {\sc O vii} emission line triplet).
Likewise, NGC~5643 X-1 was not included in the background extraction
region. We generated AGN+background and background spectra with 20000
spectral channels to optimally exploit the RGS resolving power.
Spectra were analysed with XSPECv12.6.0 in the 0.2--2~keV
energy range, using the Cash goodness-of-fit test on the unbinned
spectrum. Details on the spectral analysis procedures can be found
in Guainazzi \& Bianchi (2007).

\begin{figure*}
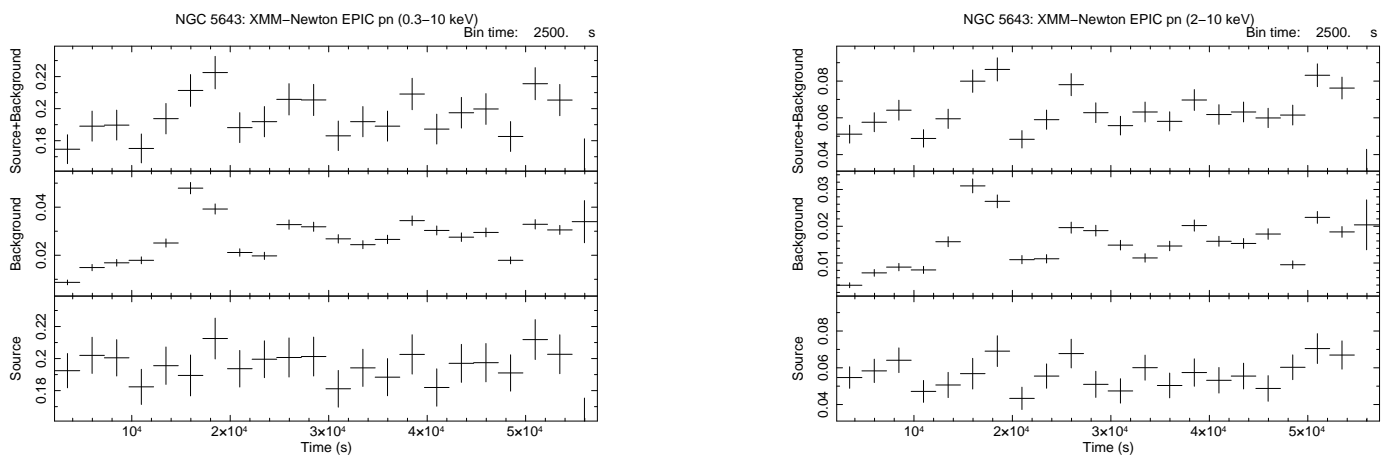

\epsfig{file=ngc5643_2_lcurve.ps,width=6cm,angle=-90}
\hfill
\epsfig{file=ngc5643_2_lcurve_hard.ps,width=6cm,angle=-90}
  \caption{0.3-10 keV (left panel) and 2-10 keV (right panel)
light curves (pn only) for the 2009 observation. The 
source+background, the background alone, and the
background-subtracted source count rates are shown in the upper, medium and lower
panels, respectively.}
  \label{lcurves}
\end{figure*}

\section{Data analysis and results on NGC 5643} 

\begin{figure*}
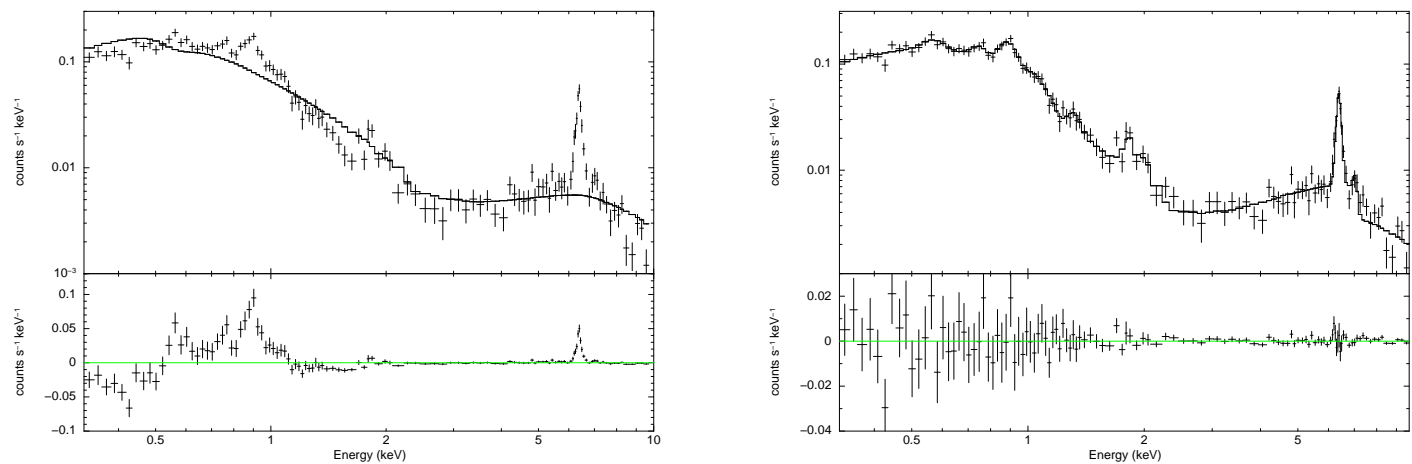

\epsfig{file=badfit.ps,width=6cm,angle=-90}
\hfill
\epsfig{file=powpex.ps,width=6cm,angle=-90}
  \caption{Left panel: spectrum and best fit model and residuals for the 2009 observation, 
when fitting
with a two power law model (pn only). Right panel: the same with the reflection model (see text),
and a narrow iron K$\alpha$ line.}
  \label{badfit}
\end{figure*}

\subsection{The 2009 observation. Phenomenological model}

In Fig.\ref{lcurves} the 0.3-10 keV (left panel) and 2-10 keV (right panel)
light curves of the pn are shown. The source+background, the background alone, and the
background-subtracted source count rates are shown in the upper, medium, and lower
panels, respectively. No statistically significant variations are apparent, 
and therefore we used the time-integrated spectrum for the spectral analysis.

In Fig~\ref{badfit} (left panel) a fit to the 0.3-10 keV pn spectrum 
with a model composed of two power laws (plus Galactic absorption,
$N_{H,Gal}=8.01\times10^{20}$ cm$^{-2}$, Kalberla et al. 2005) is presented
for illustration purposes. The fit is clearly bad, with several
features in the 0.5-1 keV band and a very prominent iron K$\alpha$ fluorescent 
line. The two power-law photon indices are about 3 (the soft one) and 
-0.87 (the hard one). The inverted hard power law suggests either significant
absorption or a reflection--dominated spectrum. 

We therefore added several narrow (i.e. unresolved)
emission lines (see Table~\ref{lines}) and 
substituted the hard power law with a pure, neutral reflection component
(see section 3.2 for a discussion of the soft X-ray emission). The soft power
law and emission lines indicate reflection from warm matter, and the neutral
reflection and the iron K$\alpha$ line from cold matter. The
fit is decent ($\chi^2$/d.o.f.=111.9/91) even if the power law illuminating the
cold reflecting matter is still unusually flat (1.11$\pm$0.21). 
Inspection of the residuals
(see Fig.~\ref{badfit}, right panel) shows that most of the remaining problems 
reside around the iron line. Indeed, an improvement is found ($\chi^2$/d.o.f.=106.7/90)
by leaving the width of the line free to vary ($\sigma$=54$^{+19}_{-24}$ eV).
A comparable fit
($\chi^2$/d.o.f.=107.1/90) is found keeping the line narrow and adding
a Compton shoulder (CS, modelled as a Gaussian with centroid energy of 6.3 keV
and $\sigma$=40 eV, see e.g. Molendi et al. 2003). A further improvement
 ($\chi^2$/d.o.f.=99.1/88) is found by adding the Ni K$\alpha$ emission line.

The CS-to-core ratio is consistent with what is expected 
in Compton-thick matter (Matt 2002). 
The iron K$\beta$/K$\alpha$ and
the Ni/Fe line ratios are also consistent (within the quite large errors)
with expectations (Molendi et al. 2003), while the 
power law indices for the best fit model are $\Gamma_S$=3.12$\pm$0.10 and
$\Gamma_H$=1.29$^{+0.08}_{-0.11}$. The hard power law is still rather flat. 
However, it must be recalled that it is not determined directly.
In fact, it is the illuminating source of the
cold reflection component (the only visible one) and,
moreover, is estimated in a limited energy band (the 
bulk of the reflection component is emitted above 10 keV).
The equivalent width of the iron K$\alpha$ line is more than 1 keV,
a value expected in a pure reflection spectrum. 

As a sanity check, we imposed the best fit model described above
to the MOS1+2 spectrum, without any fitting. 
A value of $\chi^2$/bins=99.1/118 is obtained,
demonstrating that the MOS and pn spectra are in very good agreement.

 The observed 2-10 keV flux is 
7.66($\pm0.19$)$\times$10$^{-13}$ erg cm$^{-2}$ s$^{-1}$, corresponding to a 
luminosity of 2.7$\times10^{40}$ erg s$^{-1}$.

Further tests were then applied to the pn spectrum.
The addition of a partial Compton-thin absorber in front of
the Compton reflection component does not improve the quality of the fit. 
Pure absorption instead of reflection is also
not a viable option. The fit is decent ($\chi^2$/d.o.f.=109.2/89),
but the power law index is very flat ($\Gamma_h$=0.25$^{+0.18}_{-0.35}$, with
 N$_H$=6.6$^{+1.7}_{-5.2} \times$10$^{22}$ cm$^{-2}$: if the power law index is forced to the
more reasonable value of 1.5, the fit is definitely worse, $\chi^2$/d.o.f.=129.1/90), unless
the absorption is partial ($\chi^2$/d.o.f.=111.0/89, covering fraction of 0.925$^{+0.018}_{-0.012}$).
Moreover, the very large iron line EW (about 1.5 keV with respect to the unabsorbed primary
continuum) remains unexplained. 

Therefore, from the analysis of the broad band X--ray spectrum of NGC~5643 we
can conclude that the spectrum is entirely due to reflection from warm
and cold matter, the typical spectrum of a Compton-thick source. This
conclusion is strengthened by the 2-10 keV/[O III] flux ratio. The
de-reddened [O III] flux is about 7$\times$10$^{-12}$ erg cm$^{-2}$ s$^{-1}$
(Bassani et al. 1999), and the flux ratio is therefore well within the Compton-thick regime
(Bassani et al. 1999, Lamastra et al. 2009, Marinucci et al. 2012).

\begin{table}
\small
  \caption{Centroid energies, fluxes and equivalent widths (EW) of the emission lines. EW
are calculated for the best-fit fluxes and refer to the total underlying continuum. }
  \begin{tabular}{cccc}
    \hline
& & & \\     
    & E & Flux & EW   \\
    & (keV) & (10$^{-5}$ ph cm$^{-2}$ s$^{-1}$) & (eV)  \\
& & & \\     
\hline
& & & \\  
    O {\sc vii} K$\alpha$  & 0.576$^{+0.011}_{-0.008}$  & 2.85$^{+0.54}_{-0.50}$ & 95 \\
    O {\sc viii} K$\alpha$ & 0.652$^{+0.030}_{-0.028}$ & 0.65$\pm$0.30 & 32 \\ 
    O {\sc vii} RRC/Fe {\sc xvii} 3F        & 0.740$^{+0.011}_{-0.012}$  & 1.38$\pm$0.25 & 94 \\
    O {\sc viii} RRC       & 0.874$\pm$0.009  & 1.25$\pm$0.19 & 136 \\
   Ne {\sc ix} K$\alpha$   & 0.928$^{+0.013}_{-0.015}$  & 0.78$\pm$0.17 & 105 \\
   Ne {\sc x} K$\alpha$  & 1.040$^{+0.013}_{-0.017}$  & 0.52$\pm$0.14 & 113 \\
   Fe {\sc xviii} L & 1.112$\pm$0.036  & 0.21$\pm$0.09 & 59 \\
   Mg {\sc xi} K$\alpha$ & 1.318$\pm$0.038 & 0.15$\pm$0.07  & 67  \\
   Si {\sc xiii} K$\alpha$ & 1.850$^{+0.40}_{-0.34}$  & 0.13$^{+0.09}_{-0.08}$ & 113 \\
   Si {\sc xiv} K$\alpha$  & 2.012$\pm$0.055  &  0.08$\pm$0.05 &  77 \\
   Fe K$\alpha$  & 6.403$\pm$0.011 & 1.28$^{+0.19}_{-0.16}$ & 1204 \\ 
   Fe K$\alpha$ CS & 6.3 & 0.17$^{+0.10}_{-0.16}$ & 160 \\ 
   Fe K$\beta$   & 7.06  & 0.12$\pm$0.08 & 126 \\
   Ni K$\alpha$  & 7.43$\pm$0.009 & 0.14$\pm$0.08 & 168 \\ 
& & & \\     
    \hline
  \end{tabular}{\noindent}
\label{lines}
\end{table}

\subsection{The 2009 observation. The soft X-ray emission}

The presence of lines with such different ionization potentials, along with the
presence of radiative recombination continua (RRC), argue in favour
of photoionized emitting matter. In fact, substituting in the pn spectrum
the soft power law
and the emission lines (apart from the iron and nickel lines from neutral matter)
with a thermal plasma component ({\sc mekal} model in {\sc XSPEC}), 
an unacceptable fit ($\chi^2$/d.o.f.=239.6/109) is obtained.
By adding a second thermal component the fit is still significantly worse than with
a power law plus emission lines ($\chi^2$/d.o.f.=152.9/106), and with metal abundances
very low, less than 10\% the solar value. Not even the addition of a third thermal
component suffices to obtain a good fit ($\chi^2$/d.o.f.=142.1/103).

To check whether the soft X-ray emission is indeed due to photoionized matter,
we substituted the phenomenological model (i.e., a power law plus as many emission lines
as required by the data) by a physical model based on {\sc cloudy} 
(last described in Ferland et al. 1998). Details on the model can be found in 
Bianchi et al. (2010); here, we just recall that it is a tabular model, 
with the tables built assuming solar abundances, an illuminating continuum
as described in Korista et al. (1997) and an electron density of 10$^3$ cm$^{-3}$.
The model is rather insensitive to the latter parameter over a wide range of values.

A single photoionized region is clearly insufficient ($\chi^2$/d.o.f.=256.2/107),
and the addition of a second ($\chi^2$/d.o.f.=210.7/104) and even of a third
($\chi^2$/d.o.f.=161.6/101) and a fourth ($\chi^2$/d.o.f.=143.2/98) region, 
while improving the quality of the fit,
is not sufficient to provide an acceptable fit. This contrasts with other
sources where spectra of comparable signal-to-noise are well fitted by a
small number of photoionized regions (e.g., Bianchi et al. 2010; Marinucci et al. 2011).

The bad fit is largely due to residuals around 0.75 keV and 1.8 keV,
possibly related to enhanced O {\sc vii} RRC and/or Fe {\sc xvii} 3F line, and
Si {\sc xiii} K$\alpha$ line, respectively. Indeed,
adding two emission lines in addition to three photoionized regions results in a 
much better, and now acceptable, fit ($\chi^2$/d.o.f.=107.4/97). This may correct for some rigidities
in the adopted model, such as the use of solar abundances and the assumption of
well separated emitting regions. The parameters for this fit are
summarized in Table~\ref{bestfit} (Model A). 
No further improvement is found with a fourth photoionized region.
One emitting region is thick and not very ionized, while the other two are much 
thinner and have moderate and high ionization, respectively.

Alternatively, the inability of a pure photoionization model to reproduce the
data well may be due to the presence also of collisionally ionized plasma. This hypothesis
is corroborated by the best-fit energies of the O {\sc vii} and Ne {\sc ix} triplets,
which suggest a non-neglibile contribution by the resonant lines. A decent fit
is, however, obtained only with no less than two photoionized and one collisionally ionized regions
($\chi^2$/d.o.f.=134.5/101). A better fit is obtained 
adding either a further collisionally ionized region ($\chi^2$/d.o.f.=118.2/98)
or a third photoionized region ($\chi^2$/d.o.f.=117.2/98). Adding a second collisionally
ionized region to the latter fit does not lead to any improvement. Because a feature
in the residuals is still apparent around 1.8 keV, an emission line was added, 
resulting in $\chi^2$/d.o.f.=108.1/97. The parameters for the latter fit are
summarized in Table~\ref{bestfit} (Model B).

With respect to model A, the high ionization photoionized emitting region is almost
unchanged. The low ionization region is now thin, while the moderate ionization is thick.
The metal abundance in the collisionally ionized component, $A$, is very 
poorly constrained, so it has been fixed to 1.
The emission measure is 8.6$\times$10$^{61}$ cm$^{-3}$. 
In order for the emitting region to be optically thin
to Compton scattering, its radius must be larger than about 10$^{13}$ cm, a value 
that does not give any valuable information on the nature of this region.

It must be pointed out that in both model A and B the index of the hard power law
is rather
flat, less than about 1.5 at 90\% confidence level. As remarked above, this parameter
is determined indirectly from the reflection component, which dominates only above 3 keV,
where the quality of the spectrum is modest (see Fig.~\ref{badfit}). 

\begin{figure}
\epsfig{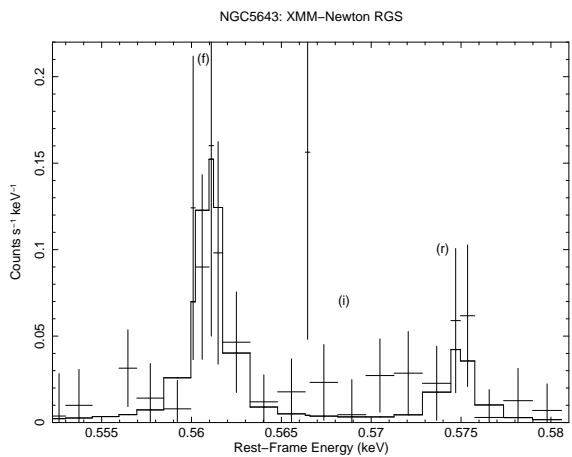}
  \caption{The RGS spectrum and model of the {\sc o vii} complex.}
  \label{triplet}
\end{figure}

We also analysed the RGS data to try to separate the line multiplets into their 
components and exploit their diagnostic power. Three emission lines have been detected:
the {\sc O vii} K$\alpha$ forbidden and resonant lines at 0.5610$\pm$0.0004 keV and 
0.575$^{+0.003}_{-0.001}$ keV, respectively, and the {\sc O viii} K$\alpha$ doublet 
(unresolved) at 0.6531$\pm$0.0006 keV. Fluxes are 11$^{+5}_{-4}\times10^{-6}$ ph cm$^{-2}$ s$^{-1}$,
3.1$^{+4.0}_{-2.6}\times10^{-6}$ ph cm$^{-2}$ s$^{-1}$, 
and 5.9$^{+2.2}_{-1.8}\times10^{-6}$  ph cm$^{-2}$ s$^{-1}$,
respectively (see Fig.~\ref{triplet} for the {\sc O vii} complex). Both the
 {\sc O vii} forbidden/resonant ratio (Porquet \& Dubau 2000) and the  
{\sc O vii}/{\sc O viii} ratio (Guainazzi \& Bianchi 2007)
suggest that the dominant process is photoionization,
confirming the results from the broad band analysis  
even if, as discussed above, a further contribution by collisionally ionized plasma
cannot be ruled out.

From the best fit parameters it is not easy to derive the location of the
ionized reflecting regions, because we do not know the density of the matter
(the fit is largely insensitive to this parameter) and the intrinsic luminosity
of the sources. Assuming a density of 10$^{3}$ cm$^{-3}$ (appropriate for the NLR),
a factor 70 between instrinsic and reflected luminosities (as appropriate for 
a Compton-thick source, Marinucci et al. 2012), and an average power law index
of 1.7 to extrapolate the 2-10 keV luminosity dwon to 1 Rydberg, we obtain
for the three reflecting zones (model B) an inner radius of about 2, 1, and 0.15 parsecs,
respectively. The outer radius depends on the radial dependence of the density, and can
be derived from the best fit values for the column density. Assuming that the higher
column-density, brighter reflector is the one responsible for 
the extended emission observed by Chandra up to 500-600 pc 
(Bianchi et al. 2006), a $r^{-1}$ dependence
of the density is found, while the other two reflectors are too small to be 
resolved even by Chandra, and even assuming a  $r^{-2}$ radial dependence.

\begin{table}
  \caption{Parameters of the best fit models of the soft X-ray emission. See
text for details. $U$ is the ionization parameter defined as $N/(4\pi R^2 c n_e)$,
where $N$ is the number of ionizing photons above 1 Ry. $\epsilon$ is the normalization
of the {\sc mekal} component.
The fluxes for each photoionized zone's emission are corrected for absorption.
All units are in cgs, apart from energies which are in keV. Line fluxes are
in photons per unit area and time.
Errors indicated with (a) are pegged to their limits. }
  \begin{tabular}{ccc}
    \hline
& &  \\     
      & Model A & Model B   \\
& &  \\  
    \hline
& &  \\  
 log$U_1$ & 0.91$^{+0.12}_{-0.17}$ & 0.30$^{+0.28}_{-0.41}$ \\     
logN$_{H,1}$ &  23.5$^{+0.0(a)}_{-0.8}$  &  20.4$^{+0.9}_{-0.9}$\\     
$F_1$ (0.5-2 keV) &  6.13($\pm0.50$)$\times10^{-14}$ & 4.16($\pm0.38$)$\times10^{-14}$ \\
 log$U_2$  &  1.33 $^{+0.06}_{-0.03}$  &  1.16$^{+0.07}_{-0.12}$ \\     
logN$_{H,2}$ & 19.7$^{+0.2}_{-0.4}$   &  22.4$^{+0.3}_{-0.1}$ \\     
$F_2$ (0.5-2 keV) & 1.04($\pm0.49$)$\times10^{-13}$   &  7.66($\pm0.52$)$\times10^{-14}$ \\
 log$U_3$ & 2.57$^{+0.08}_{-0.09}$  & 2.59$^{+0.07}_{-0.07}$ \\     
logN$_{H,3}$ & 19.4$^{+0.7}_{-0.4}$  & 19.7$^{+0.1}_{-0.1}$ \\     
$F_3$ (0.5-2 keV)  &  4.16($\pm0.42$)$\times10^{-14}$  & 4.55($\pm0.44$)$\times10^{-14}$ \\
$kT$  & -- &  0.53$^{+0.10}_{-0.09}$ keV \\
$A$  & -- & 1 (fixed) \\    
$\epsilon$  & -- & 2.49$^{+0.54}_{-0.45}\times10^{-5}$ \\    
E$_1$ & 0.746$^{+0.011}_{-0.009}$ & -- \\
F$_1$ & 1.41$^{+0.21}_{-0.24}\times10^{-5}$ & -- \\
E$_2$ & 1.84$^{+0.04}_{-0.04}$  & 1.84$^{+0.03}_{-0.05}$ \\
F$_2$ & 9.8$^{+8.7}_{-5.2}\times10^{-7}$ &  1.24$^{+0.85}_{-0.79}\times10^{-6}$ \\
$\Gamma$ & 1.26$^{+0.25}_{-0.33}$ & 1.23$^{+0.26}_{-0.35}$ \\
& &  \\  
\hline
  \end{tabular}{\noindent}
\label{bestfit}
\end{table}

\subsection{The 2003 observation}

\begin{figure}
\epsfig{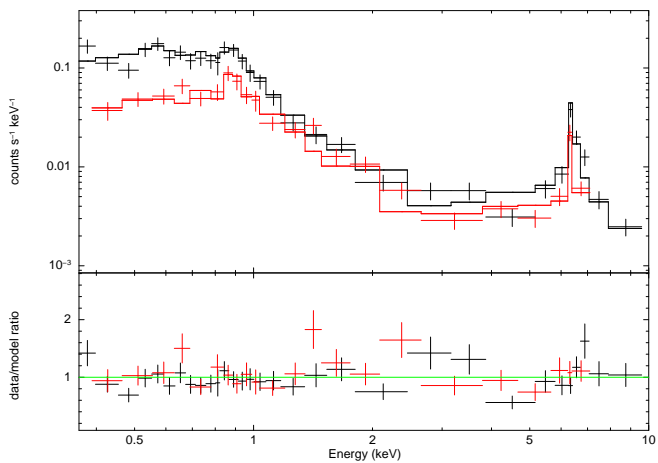}
  \caption{The 2003 spectrum with superimposed the best fit model for
the 2009 spectrum. Both pn {\bf (black)} and MOS1+2 {\bf (red)} spectra are shown.}
  \label{bestfit_obs1}
\end{figure}

We then imposed the best-fit phenomenological model described
above to the spectra (both pn and MOS1+2) 
of the 2003 observation without any fitting or renormalization
procedure. The result is shown in Fig.~\ref{bestfit_obs1}.  
By leaving the reflection component and the iron line parameters 
free to vary, parameters values consistent with
those found in the 2009 observation are obtained.
We therefore conclude that there is no significant 
variation between the two observtions.

\section{Other sources in the field}

\begin{figure}
\epsfig{file=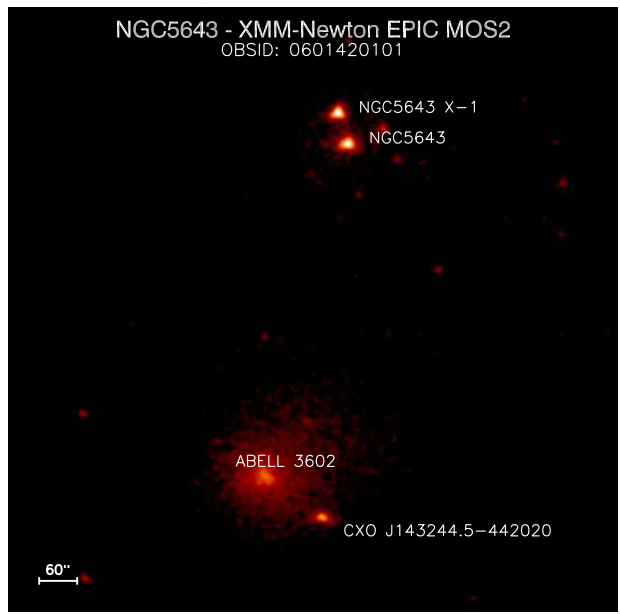,width=8cm}
  \caption{The MOS2 field of view.}
  \label{image}
\end{figure}

There are other bright sources in the field of NGC5643 (see Fig.~\ref{image}), the 
most prominent one being NGC5643 X-1, which is very close to the active nucleus
and one of the brightest ULX known (Guainazzi et al. 2004). There is also
an extended source, to be associated with the cluster of galaxies Abell 3602,
and a point-like source, possibly a galaxy belonging to Abell 3602,
and identified with the CXOJ143244.5-442020 Chandra source. 
The coordinates of all these sources are listed in Table~\ref{sources}. 

In all fits described below, the same 
Galactic absorption as for NGC~5643 is assumed.

\begin{table}
  \caption{The coordinates of the brighest sources in
the field of view.}
  \begin{tabular}{ccc}
    \hline
 & & \\     
     Source & R.A. (J2000)  & Dec (J2000)  \\
& &  \\     
\hline
& & \\  
NGC 5643 (Nucleus) & 14 32 40.7 & -44 10 27.9 \\
NGC 5643 X-1 & 14 32 42.3 & -44 09 39.2  \\
Abell 3602 & 14 32 53.2 & -44 19 19.8 \\
CXOJ143244.5-442020 & 14 32 44.5 & -44 20 20.3 \\
& & \\     
    \hline
  \end{tabular}{\noindent}
\label{sources}
\end{table}

\subsection{NGC~5643 X-1}

\begin{figure}
\epsfig{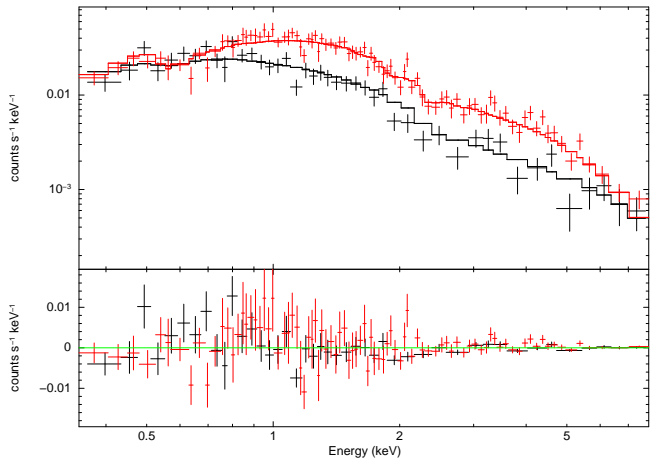}
  \caption{EPIC Spectra and best fit model and residuals for NGC 5643 X-1 (black: pn, red: MOS).}
  \label{X1}
\end{figure}

Following Guainazzi et al. (2004), we fitted the pn and MOS1+2 spectra with a 
model consisting of a power law absorbed by both Galactic and local materials. 
The intrinsic absorber is much smaller ($N_H <$3$\times$10$^{20}$ cm$^{-2}$)
than in 2003, suggesting a variation in the local environment. In the subsequent
fits, only Galactic absorption was included. The
spectra are rather noisy, also because the source is partly in a gap in the pn, 
with a significant reduction in detected counts. The fit is not very good
 ($\chi^2$/d.o.f.=166.1/133), but no clear features are 
present in the residuals (see Fig.~\ref{X1}). The power law index
is 1.65$\pm$0.04, which is consistent with the 2003 value. The 2-10 keV flux is 
3.52($\pm0.10$)$\times$10$^{-13}$ erg cm$^{-2}$ s$^{-1}$, more than a factor of two
lower than in the previous observation, thus indicating a change in the 
luminosity but not in the intrinsic shape of the emission. The corresponding
luminosity is 1.2$\times10^{40}$ erg s$^{-1}$ , if the source belongs to NGC 5643.

Adding a multicolour black body disk component ({\sc diskbb} in {\sc xspec}), 
as is customary for this class of sources, a 
slightly better fit is found ($\chi^2$/d.o.f.=156.0/131, improvement significant at the
98.4\% confidence level according to the F-test). The power law index
is now 1.42$\pm$0.13 and the inner disk temperature is 0.37$\pm$0.06 keV. 
The inner radius is about 790/$\sqrt{cos\theta}$ km ($\theta$ being the disk
inclination angle). Assuming that the inner radius coincides with the ISCO
for a Schwarzschild black hole, a black hole mass of about 90/$\sqrt{cos\theta}$
solar masses would be derived. Even higher masses would be, of course, derived for
a spinning black hole. 

\subsection{Abell 3602}

This source is clearly extended (see Fig.~\ref{image}), and should be identified
with the cluster of galaxies Abell 3602, even if the 
NED coordinates (taken from Abell et al. 1989) are slightly different 
than ours and seem to refer instead to CXOJ143244.5-442020 (see next paragraph). 

Abell 3602 was observed by the $Einstein$ observatory and has an entry in the ROSAT 
Bright Source catalogue. No hard X-ray data have been published yet to our knowledge. Its
optical extension is about 150''$\times$150'' (Abell et al. 1989). The redshift
is estimated to be 0.1044 (White et al. 1997, adopting the tenth brightest galaxy distance 
estimator) or 0.1047 (photometric redshift, Coziol et al. 2009).

\begin{figure}
\epsfig{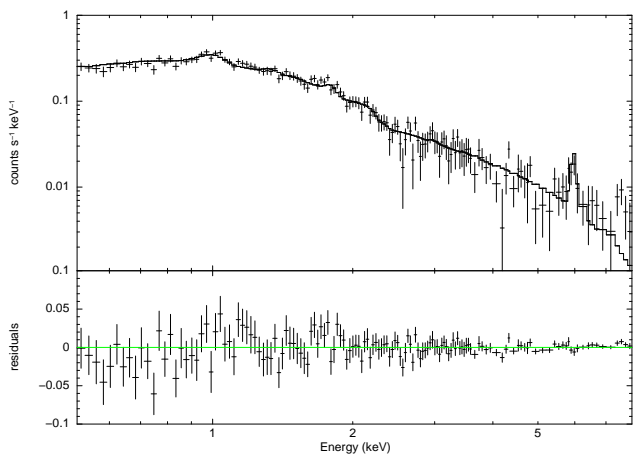}
  \caption{pn spectrum and best fit model and residuals for Abell 3602.}
  \label{abell}
\end{figure}

The spectra were extracted in a region of 90'' in radius to exclude the point-like source
CXOJ143244.5-442020 (see Fig.~\ref{image}). At the redshift of the cluster
this angular radius corresponds to 173 kpc. 
The spectra of 2003 and 2009 are perfectly consistent with each other. For the
spectral fitting we only used the 2009 data, because the addition of the 2003
spectrum does not increase the S/Nsignificantly. 
The pn spectrum is well fitted ($\chi^2$/d.o.f.=138.2/140, see Fig.~\ref{abell}) 
by an optically thin thermal
plasma ({\sc mekal} model in {\sc xspec}) with a temperature
of 4.37$^{+0.28}_{-0.29}$ keV\footnote{A mass $M_{500}\sim$3.2$\times$10$^{14}$ M$_{\odot}$
can be derived from the M-T relationship of Arnaud et al. (2005).}, a metal abundance (with respect to
solar) of 0.49$^{+0.12}_{-0.11}$, and a redshift of
0.120$\pm$0.008 (not too different from the optical values). 
The 2-10 keV flux is 1.52($\pm0.05$)$\times10^{-12}$ erg cm$^{-2}$ s$^{-1}$, and the 
resulting 2-10 keV luminosity is 4.4$\times10^{43}$ erg s$^{-1}$, using the
optical redshift.

\subsection{CXOJ143244.5-442020}

Not very distant from the center of Abell 3602 
(137'', corresponding to about 260 kpc at the cluster's optical 
redshift), there is another
bright X-ray source, CXOJ143244.5-442020. 
The 2009 spectrum of the source is well fitted ($\chi^2$/d.o.f.=60.8/58)  
by a single power law with $\Gamma$=2.02$\pm$0.07. The 2-10 keV flux
is 1.90($\pm0.23$)$\times10^{-13}$ erg cm$^{-2}$ s$^{-1}$. If the source belongs
to Abell 3602, this corresponds to a 2-10 keV luminosity of 5.3$\times10^{42}$ erg s$^{-1}$.

When adding a narrow neutral iron K$\alpha$ line at 6.4 keV  and leaving 
the source redshift free to vary, only a slight improvement in the fit is 
found ($\chi^2$/d.o.f.=56.3/56, significant at the 88\% confidence level
according to the F-test).
Interestingly, the redshift is constrained to be in the range [0.08, 0.17] which, albeit
with a large error, is consistent with that of Abell 3602 (see previous
paragraph). Fixing its value to that of the cluster, the iron line equivalent width
is 0.35$\pm$0.29 keV. 

The 2003 spectrum is very poor. Not surprisingly, a power law fit provides
again a good fit ($\chi^2$/d.o.f.=7.7/13),  with $\Gamma$=2.22$\pm$0.15. The 2-10 keV flux
is 2.10($\pm0.5$)$\times10^{-13}$ erg cm$^{-2}$ s$^{-1}$. 

The most likely hypothesis is that this source is an unobscured AGN,
possibly belonging to Abell 3602.

\section{Summary}

The relatively long XMM-$Newton$ observation of NGC~5643
performed in 2009 has solved 
the two main issues left open by the previous, shorter observation,
performed in 2003. In fact, it has been possible to  
show that the source is Compton-thick. Regarding 
the nature of the soft X-ray emission, the 
dominant contribution is by photoionized matter, even if it is still 
unclear whether collisionally ionized matter contributes as well. 

The ULX NGC 5643 X-1 was confirmed to be very luminous, even if more 
than a factor 2 fainter than in 2003. 

Two other bright sources are also present in the field if NGC 5643.
The cluster of galaxies Abell 3602 is one of them, and we are able to
provide the first good quality spectrum above 2 keV.  The second source,
CXOJ143244.5-442020, is likely an unobscured AGN belonging to Abell 3602.

\begin{acknowledgements}
We thank the referee, Marco Salvati, for useful comments which helped to
improve the clairty of the paper, and 
Stefano Ettori for useful discussions on Abell 3602.
Based on observations obtained with XMM--$Newton$, an ESA science mission with 
instruments and contributions directly funded by ESA Member States and the USA
(NASA). GM and SB acknowledge financial support from ASI under grant I/009/10/0.
\end{acknowledgements}

\end{document}